\documentclass[journal]{IEEEtran}
\usepackage{amsmath}
\usepackage{graphicx}
\usepackage{listings} 
\usepackage{tabularx}

\graphicspath{{../figs}}

\ifCLASSINFOpdf
\else
\fi
\begin{document}
%
\title{Subjective Assessment Experiments That \\ Recruit Few Observers With Repetitions (FOWR)}
%
%
%

\author{Pablo~P{\'e}rez,
        Lucjan~Janowski,
        Narciso~Garc{\'i}a,
        and~Margaret~Pinson
\thanks{P. P{\'e}rez is with Nokia Bell Labs. L. Janowski is with AGH University of Science and Technology. N. Garc{\'i}a is with Universidad Polit{\'e}cnica de Madrid. M. Pinson is with the Institute for Telecommunication Sciences (ITS).}
\thanks{The work done by P. P{\'e}rez and N. Garc{\'i}a has received funding from the European Union’s Horizon 2020 research and innovation programme under grant agreement no 957102 (5G-RECORDS). 
The work done by N. Garc{\'i}a has also been partially supported by the Ministerio de Ciencia, 
Innovación y Universidades (AEI/FEDER) of the Spanish Government under 
project TEC2016-75981 (IVME). The work done by L. Janowski has received funding from the Norwegian Financial Mechanism 2014-2021 under project 2019/34/H/ST6/00599.}%
\thanks{Manuscript received XXXX XX, XXXX; revised XXXX XX, XXXX.}}

%
%

\markboth{IEEE Transactions on Multimedia,~Vol.~XX, No.~XX, Month~Year}%
{Perez \MakeLowercase{\textit{et al.}}: FOWR}
%

\IEEEpubid{0000--0000/00\$00.00~\copyright~2015 IEEE}


\maketitle

\begin{abstract}
Recent studies have shown that it is possible to characterize subject bias and variance in subjective assessment tests. Apparent differences among subjects can, for the most part, be explained by random factors. Building on that theory, we propose a subjective test design where four to six team members each rate the stimuli multiple times. The results are comparable to a high performing objective metric. This provides a quick and simple way to analyze new technologies and perform pre-tests for subjective assessment.
\end{abstract}


%
\IEEEpeerreviewmaketitle


\section{Introduction}

Industry depends on video, image, and audio quality assessments when making important business decisions. Examples include optimizing video encoders, choosing video transmission bandwidths, fine tuning networks, and agreeing upon standard speech codecs for future cellular networks. 
The traditional options are to either conduct a subjective test or choose a well-established objective metric. 

Both options have advantages and limitations \cite{staelens2015measuring}.
Subjective tests are accurate but time-consuming. 
Objective metrics are inaccurate but fast. 
Moreover, metrics quickly become unreliable when applied to new technologies or scenarios outside of their intended scope. For example, a metric designed for broadcast video applications may yield random ratings when given user-generated content. 
Neither option meets the needs of a company to quickly assess new technologies during the research and development process. 

This paper proposes a compromise solution: a subjective test where four to six team members rate each stimulus multiple times. 
This would reduce costs associated with recruiting and handling subjects, shifting the work onto team members who are obliged to assist. 
Obvious quality impairments are easily spotted by all subjects (particularly experts), and novel technologies are not problematic.  
This compromise would provide a viable third option, as long as the accuracy lies somewhere between the accuracy of subjective test data and objective metric values.  
We have named the resulting test protocol FOWR: Few Observers With Repetitions.

In this paper, we will review previous studies of subject rating behaviors, which provide the theoretical basis for this proposal. 
We then present a subjective test that was designed to evaluate this FOWR experiment design.
The conclusions reached by this test will be compared with conclusions reached by a test conducted with the traditional experiment design. 

\subsection{Related Work}
\IEEEpubidadjcol
The conventional experiment design is well described in the literature. In literature not related to video or audio quality a typical subjective experiment is called a within-subjects design. Details about analysis of such experiments can be found in classical literature from sociology like chapter 14 of \cite{ExperimentDesign}.

We cannot take short-cuts when choosing stimuli. An illustrative example is Fig. 3 of \cite{pinson2013selecting}. Thirteen experiments conducted at various labs investigated the same issue: an equation that maps audio quality and video quality (measured separately) to the overall audiovisual quality (assessed jointly). Eight experiments used one or two source videos and reached disparate conclusions; five experiments used five to ten source videos and reached similar conclusions.
Subjective and objective analyses are only reliable if care is taken when choosing media stimuli and impairments. 

However, the number and nature of subjects are more open to discussion. 
The decision about the number of subjects should be based on so-called power analysis \cite{powerAnalysis}. This analysis is simple for classical problems where the effect size is well understood, but in the case of new technologies and less classical experiments it is difficult to say what is the effect size. In this case a pilot study is a good option, but we have to carefully analyze the pilot study data. A low number of participants can have a strong influence on the obtained results \cite{brysbaert_2018}.

An understanding of the specificity of subjective experiments is key for correctly planning experiments. 
One issue is to analyze how the number of subjects influences the obtained results.
 Pinson et al. \cite{pinson2012influence} presents a detail analysis of the influence of number of subjects and the environment. 
The main conclusion was that 24 subjects is a reasonable number. A different conclusion was drawn by Winkler in \cite{OnTheProperties_Winkler_2009}, where 15 was recommended. 
The difference could be the result of complicated behaviors related to the scoring process, as described in \cite{Digital_Akramullah_2014} pages 118--119.

Pinson \cite{SubjObjCI} analyzes ratings from 60 subjective tests to estimate confidence intervals (CI) of mean opinion scores (MOS) and the likelihood that two labs will disagree (i.e., \textbf{A} has significantly better quality than \textbf{B}, or vice versa).
Assuming a well-designed and carefully conducted subjective test using the 5-level absolute category rating (ACR) method, the MOS CI is approximately 0.5 for 24 subjects, 0.7 for 15 subjects, 1.1 for 9 subjects, and 1.5 for 6 subjects. 
Unknown factors can lead to higher CIs, but usually no more than the next category (e.g., a 24 subject test with CI = 0.7).
The likelihood that two subjective test labs will disagree on the rank ordering of stimuli is $\approx0.17\%$ with a long tail extending up to $1\%$ and an outlier at $1.84\%$. These lab-to-lab comparisons show similar rating behaviors for speech, image, and video quality experiments.

Additionally, Brunnstr\"om and Barkowsky \cite{brunnstrom2018statistical} analyze how the discriminating power of experiments decreases when the number of comparisons increases. 
A typical subjective experiment with 24 subjects and  100 pairwise comparisons (about 10 points compared pairwise) cannot discriminate reliably between levels smaller than 1.0 MOS differences in a 1-5 scale.
The sample size required to discriminate between different levels for 0.5 MOS differences increases to 80 subjects.

A few studies have compared expert and non-expert rating behaviors. 
Speranza et al. \cite{5518177} found a small but statistically significant difference between expert MOSs and non-expert MOSs, particularly at low bit-rates where experts were more critical than non-experts. Note that the authors assume that subjective tests produce absolute MOSs, which was later refuted \cite{pinson2012influence}. Speranza et al. did not compare the relative ranking of stimuli MOSs, but a visual examination of the \cite{5518177} scatter plots indicates a simple relationship: experts used more of the 100-level scale and had a negative bias. 

Kumcu et al. \cite{7781646} compared expert and non-expert ratings for two different applications: medical imagery and denoising. Their experts and non-experts produced comparable scores except for the rank ordering of two denoising systems. Recent analyses of \cite{7781646} and similar datasets indicate that experts and non-experts agree on the rank ordering of stimuli quality, when statistical equivalence is taken into consideration (revision to \cite{SubjObjCI}, pending publication). However, experts may have an increased or decreased sensitivity to certain impairments. 
Thus, for some applications, experts and non-experts will act like two pools of subjects selected at random from the same population, while for other applications, experts may reach different conclusions than non-experts about whether MOSs are equivalent or significantly different.

Several researchers have tried to describe and model the voting processes. 
Streijl, Winkler, and Hands \cite{Streijl2016} present a detail analysis of the 
limitations of Mean Opinion Scores (MOS).
Mean opinion scores are not a precise number, they note, but a statistical measurement with some uncertainty.
This statistical nature has the advantage of allowing the use of parametric statistics on the MOS analysis, as the averaging of subject scores shows normal behavior \cite{narwaria2018data}. 
However, mean values might not capture all the relevant information of subjective scores, and some alternatives have been proposed, such as 10\% or 90\% quantiles \cite{hossfeld2017no},
discrimination between pairs or similar or dissimilar stimuli  \cite{krasula2016accuracy}, or quality score distributions \cite{wu2011learning,seufert2019fundamental}.

Recent analyses try to build a mathematical model that describes subject rating behaviors. Janowski and Pinson \cite{janowski2015accuracy} model subject voting as a stochastic Gaussian process, with users presenting a bias and randomness that deviate from the target ground truth. The ground truth is also affected by a random noise associated to the difficulty of rating each Processed Video Sequence (PVS).
Variations of this model have been proposed for processing of noisy subjective data \cite{li2017recover}, modeling  just-noticeable-difference scores \cite{wang2018user}, or analyzing Adaptive Media Playout quality \cite{perez2019subjective}.

The model proposed in \cite{janowski2015accuracy} and described by notation presented in \cite{Janowski2019NotationFS} is defined by

\begin{equation} \label{eqn:subjectmodel}
U_{ijr} = \psi_{j} + \Delta_{i} + \upsilon_{i} X + \phi_{j} Y
\end{equation}

Where 

\begin{itemize}
	\item $U_{ijr}$ is a random variable related to a vote
	\item  $u_{ijr}$ is the observed rating for subject $i$, PVS $j$, and repetition $r$ 
	\item $\psi_j$ is the true quality of PVS $j$ (i.e., ground truth)
	\item $\Delta_i$ is the subject bias (i.e., overall shift between the $i$th subject's ratings and the true quality)
	\item $\upsilon_i$ is the magnitude of the $i$th subject's random noise 
	\item $\phi_j$ is the magnitude of the $j$th PVS's random noise
	\item $X$ and $Y$ are random variables with normal distribution $\mathcal{N}(0,1)$
\end{itemize}

Janowski and Pinson show that subject bias is quite stable  \cite{janowski2015accuracy}. 
The observed subject biases, $\Delta_i$, follow a normal distribution:

\begin{equation}
\label{eqn:bias}
\Delta_i \sim \mathcal{N}(0, \sigma_{\Delta})
\end{equation}
where they observe $\sigma_{\Delta}=0.34$. They also observe that the combined distributions of $\Delta_i$, $\upsilon_i$ and $\phi_j$ span about $\pm 25\%$ of the rating scale, and the ranking of sequences is quite stable across subjects. 
Of particular importance to this paper, the model underlying \eqref{eqn:subjectmodel} indicates that the number of subjects in an experiment can be reduced if each subject scores each PVS multiple times:
by repeating the experiment, the mean score of each individual subject for each individual PVS should converge to the expected value of $U_{ijr}$, which represents the \textit{true opinion} of the subject $\bar{u}_{i,j}$:

\begin{equation}
\label{eqn:true_opinion} 
\bar{u}_{i,j} = E\left[U_{ijr}\right] = \psi_j + \Delta_i 
\end{equation}

Subject bias can change slightly over the duration of a long experiment or in response to certain subject matter. 
For example, dataset ITS4S2 \cite{its4s2} is divided into 14 sessions, each with photographs on a different topic (e.g., landscapes or disasters). Visual inspection of subject biases indicates that only one of 16 subjects had notable session biases: liking tourist photos and disliking photos of places. 
Janowski and Pinson \cite{janowski2015accuracy} measure statistically significant differences in subject bias between sessions for 6 of 26 subjects in \cite{aghntia} but conclude that these perturbations are small enough to be safely ignored.

\subsection{Research Questions}

Given the following findings of previous research:
\begin{itemize}
\item Subjects show a consistent additive bias.
\item  Subjects tend to agree on pairwise orderings.
\item  The number of subjects can be reduced if each subject scores each PVS several times.
\item  A well-designed and carefully conducted subjective experiment will have a resolution no better than 0.5 MOS.
\end{itemize}

We pose the following research questions:

\begin{itemize}
\item Is it possible to design a valid experiment using few observers with repetitions (FOWR) with resolution of 0.5 MOS?
\item Can these subjects be experts?
\item How can subject bias be correctly handled?
\item Would this approach be at least as reliable as objective quality scores?
\end{itemize}


\section{Experimental design}
\subsection{Subjective experiment}

To answer these questions, we conducted a subjective experiment, ITERO\footnote{ITERO full dataset, including PVSs and subjective scores, can be found at the Consumer Digital Video Library (https://cdvl.org).}, using PVSs drawn from a prior experiment, ITS4S. 
The ITS4S dataset \cite{pinson2018its4s} contains 4-second PVSs without repetition, organized into eight sessions of 100 PVSs.
For ITERO we chose the ``Everglades'' session and added 10 sequences from the ``Sports'' session, for a total of 110 PVSs.
ITS4S provides subject ratings for these sequences from two independent labs: ITS and AGH.

The aim of ITERO is to assess whether, by repeating the experiment several times, it is possible to replicate the results from ITS4S ``Everglades'' with just one or a few subjects. 
However, \eqref{eqn:true_opinion} shows that this will only be possible if we can estimate the subject bias, $\Delta_{i}$.
Our proposed solution is adding a reference: 10 sequences from a prior test, with known MOSs from 24 subjects. 
We hypothesize that we can estimate subject bias using the prior test subset and remove it from the new test's ratings.
Therefore, ITERO has 10 sequences from the ``Sports'' session (the reference) and 100 sequences from the ``Everglades'' session (the new test).

The ITERO experiment was conducted in three different labs: Nokia, AGH, and UPM.
At Nokia and AGH, each person rated sequences on their own laptop.
At UPM, the same computer was used by all the people in the experiment.
Each subject was instructed to repeat the experiment 10 times, preferably not twice on the same day.
Twenty-seven subjects took part on the ITERO experiment.
Of these, 20 finished all 10 repetitions.

\begin{table}
\caption{Post-session questionnaire}
\label{tab:questionnaire}
\begin{tabularx}{\columnwidth}{|lX|}
\hline
Item & Question \\
\hline
Confidence & It was easy to have an opinion about each sequence \\
Focus & I have been focused on the task for the whole duration of the test \\
Tiredness & I am tired of doing this test \\
\hline
\end{tabularx}
\end{table}

In each experiment session, the subject performed a screen test, rated the same 110 PVSs, and answered a questionnaire.
The screen test assessed interactions between the subjects, monitor, and environment (e.g., the subject's ability to perceive small luma differences on this monitor in the current lighting condition).
This tool has been successfully used in crowdsourcing experiments in the past \cite{hossfeld2014best,gardlo2014crowdsourcing}.
A short questionnaire was conducted after each session to assess the subject's tiredness (Table \ref{tab:questionnaire}).
Questions were answered in a 5-point Likert scale, where 5 is \textit{totally agree} and 1 is \textit{totally disagree}.

This subjective test adheres to ITU-T Rec. P.913, which ensures the rights and welfare of the human subjects (see Clause 11.1). Photographs of the subjective testing environments are omitted, due to the large diversity of locations.

\subsection{Structure of the data set}

The data from one ITERO session and a single subject will be referred to as a \textit{repetition}.
The objective of the test is to discover whether a few (e.g. 3-4) subjects of the ITERO test can, after a few repetitions, obtain similar results as in a ``traditional'' subjective experiment, conducted with more than 20 subjects.
With this aim, from the data obtained in the ITERO dataset, two subsets have been extracted: ITERO-TEN and ITERO-ONE.

ITERO-TEN includes the data from the 10 repetitions of the 20 subjects that concluded them all.
In the subsequent analysis, $u_{i,j,r}$ will represent the rating for subject $i$, PVS $j$ and repetition $r$ within ITERO-TEN dataset.

ITERO-ONE takes the first repetition of each ITERO subject.
It includes data from all 27 subjects, regardless of how many repetitions the subject completed.
Subjects were screened for outlier rejection according to Rec. ITU-R BT.500, and no one was rejected.
ITERO-ONE is itself a ``traditional'' subjective assessment experiment, with 27 different subjects scoring the PVSs for the first time.
It will be referred to as the \textit{baseline}.
The notation $\xi_j$ will be used to represent ITERO-ONE MOS value for PVS $j$.

Additionally, we will consider the original ITS4S experiment.
The ITS4S data will be referred to as the \textit{ground truth}, since ITS4S was conducted according to the best known practices in ITU-T Rec. P.913.
Consequently, the notation $\hat{\psi}_j$ will be used to represent ITS4S MOS value for PVS $j$.

\subsection{Dataset comparison}

The target of the study is to determine whether a few subjects from ITERO-TEN can, after some repetitions, be equivalent to a ``traditional'' subjective experiment or to a state-of-the art objective metric.
This equivalence will be analyzed in terms of association, agreement, perceptual similarity, and confusion analysis. 

\textbf{Association} measures the (potentially linear) correlation between two variables, which are related but may have been measured from two different populations (or two different variables of the same observed population) \cite{krippendorff1987association}.
We will measure association by computing Pearson's linear correlation coefficient (PCC), as
it is the most frequently used comparison metric both for objective and subjective scores \cite{Streijl2016}.

\textbf{Agreement} is based on the sameness or difference between two values that measure the same underlying variable \cite{krippendorff1987association}.
In our case, it would mean that ITERO-TEN scores measure the very same quantity as the \textit{baseline} or the \textit{ground truth}.
Several options are proposed in the literature, mostly designed for the evaluation of objectives scores (see e.g. ITU-T J.149).
We will use the Root Mean Square Error (RMSE), which is probably the most popular one, and is typically reported as a figure of merit in the scientific literature.
Measuring agreement is particularly relevant in our experiment due to the effect of bias: a theoretical subject \textit{true opinion} $\bar{u}_{i,j}$ would have perfect association (PCC=1), 
but not perfect agreement (RMSE=$\Delta_i$).

\textbf{Perceptual similarity} is based on the idea that individual subjects are unable to perceive small differences of quality between similar PVSs.
In particular, increasing the resolution of the measurement scale beyond the recommended 5 levels does not increase the accuracy \cite{huynh2010study}.
This finding, together with the limitations of the discriminating power of existing experiments already described in the introduction \cite{brunnstrom2018statistical},
suggest that the actual resolution of subjective scores must lie somewhere between 0.5 and 1.0 MOS points.
Considering this, we will assume that the same PVS has been rated similarly in two different experiments if its MOS rating differs by less than 0.5 points.
We will measure perceptual similarity by computing the ratio of PVSs that have been rated similarly (\textit{MOS05}).

\textbf{Confusion analysis} compares the conclusions reached by the ground truth test and the ITERO-TEN dataset. Pinson \cite{SubjObjCI} provides expected variances, based on lab-to-lab comparisons. 
We measure the differences between the conclusions reached by the ground truth and the ITERO experiment design, to understand whether the differences fall within the expected behavior of subjective testing. 

\subsection{Benchmark}
\label{sec:benchmark}

It is not likely that any subjective or objective score is going to replicate exactly the \textit{true quality} so that PCC = 1, RMSE = 0 or MOS05 = 1.
In practice, the scores extracted from ITERO-TEN should be able to match the performance of state-of-the-art objective metrics,
or the results of two different laboratories conducting the same subjective experiment.

Table \ref{tab:objective-vqeghd3} shows the performance of a few Full-Reference quality metrics with respect to the dataset VQEG-HDTV-3 \cite{vqeghdtv}.
PSNR is a traditional benchmark for objective scores.
VQuad-HD (ITU-T J.341), and VMAF \cite{li2016toward} can be considered state-of-the-art metrics for compression and, in some cases, packet loss artifacts.
PSNR and VQuad-HD objective scores have been extracted from \cite{vqeghdtv}, which includes 3-degree polynomial fitting towards the subjective score.
VMAF scores were obtained using the subset of VQEG-HDTV-3 that contains only compression artifacts \cite{li2016toward}.
Due to the different ways in which those metrics were developed, Table \ref{tab:objective-vqeghd3} is not a fair comparison among them.
Nonetheless, it provides a view on what to expect from an objective metric working \textit{within its comfort zone}.

\begin{table}
\centering
\caption{Performance of Full-Reference Metrics on VQEG HDTV3}
\label{tab:objective-vqeghd3}
\begin{tabular}{|c|c|c|c|}
\hline
Metric & PCC & RMSE & MOS05 \\
\hline
PSNR & 0.851 & 0.586 &  0.604 \\
VQuad-HD (ITU-T J.341) &   0.917 & 0.446 & 0.714  \\
VQM (ITU-T J.144) & 0.794 & 0.690 & 0.597 \\
VMAF (subset) &  0.927 & 0.390 & 0.817 \\
\hline
\end{tabular}
\end{table}

Table \ref{tab:subjective-compare} shows the same analysis, using subjective data instead of objective data. 
It performs lab-to-lab comparisons using experiments that were conducted by six or more international labs:
the common set from VQEG-HDTV \cite{vqeghdtv} and VQEG-MM2 \cite{pinson2012influence}, \cite{6603199}.
We made lab-to-lab pairwise comparisons and then picked the median over all pairs.
Table \ref{tab:subjective-compare} also compares the ITS and AGH lab data from ITS4S session ``Everglades," as these sequences constitute most of the content in ITERO.

\begin{table}
\centering
\caption{Pairwise comparison of subjective experiments (median)}
\label{tab:subjective-compare}
\begin{tabular}{|c|c|c|c|}
\hline
Experiment & PCC & RMSE & MOS05 \\
\hline
VQEG-HDTV \cite{vqeghdtv} & 0.969 & 0.434  & 0.708 \\
VQEG-MM2 \cite{pinson2012influence} & 0.965 & 0.350 & 0.833 \\
ITS4S Everglades \cite{pinson2018its4s} & 0.960 & 0.298  & 0.930 \\
\hline
\end{tabular}
\end{table}

\section{Results for individual subjects}

\subsection{Subject responses}

\begin{figure}[htb]
\centering
\includegraphics[width=0.95\columnwidth]{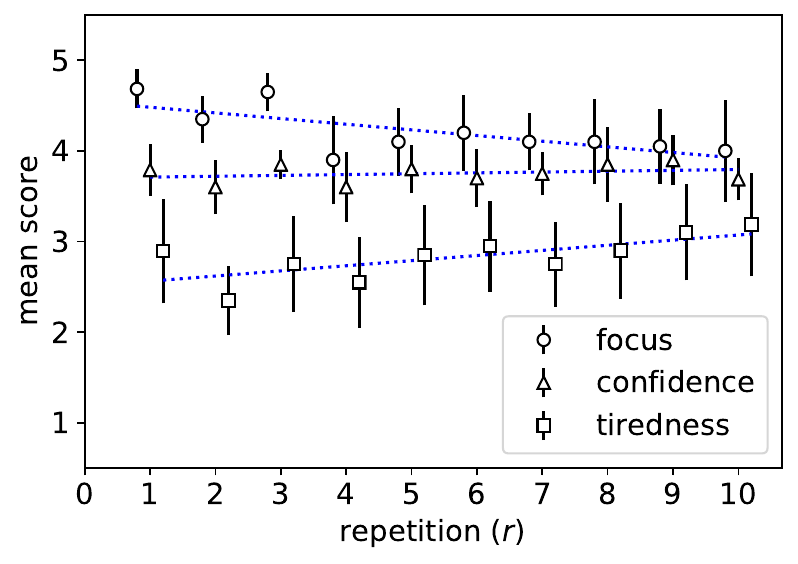} 
\caption{\label{fig:postq} Mean and 95\% confidence intervals of post-session questions along repetitions, for the 27 subjects in ITERO-TEN. 
Dashed lines show linear regressions.}
\end{figure}

The ITERO experiment was self-paced, both within each repetition and in the scheduling of repetitions on different days. 
The distribution of experiment duration was very heterogeneous: some subjects did the 10 repetitions in 12 days, while others took 8 months.
The median time between two consecutive repetitions of the same subject was 2 days.
However, in 10\% of the cases, consecutive repetitions were longer than 2 weeks apart.
Self-reported values of confidence, focus, and tiredness had slight variations in response to the repetition number (Fig.~\ref{fig:postq}).

The screen test performed at the beginning of each repetition provides a \textit{reliability index} on a 0-100 scale, which was at least 95 in 93\% of the sessions.
The sessions for which the reliability index was not at least 95 were randomly distributed across different subjects and repetitions.
We found no obvious behavior pattern that suggests that any subject or repetition should be discarded or analyzed separately.

\begin{figure}[htb]
\centering
\includegraphics[width=0.95\columnwidth]{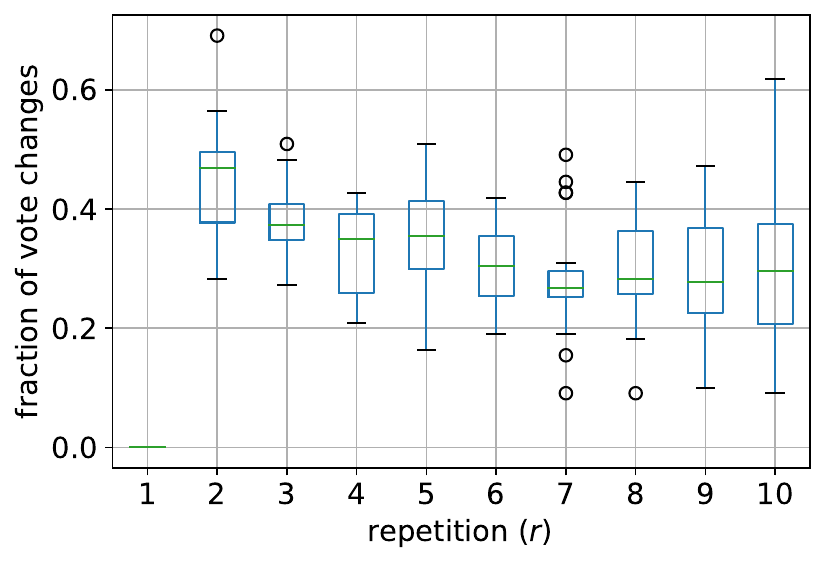} 
\caption{\label{fig:user_changes} Distribution of fraction of vote changes in one repetition with respect to the previous one, for ITERO-TEN set.
Box plots show the distribution among the 20 subjects.}
\end{figure}

Fig.~\ref{fig:user_changes} shows the fraction of changes in scoring for the same PVS in consecutive sessions, for the 20 subjects that did all 10 repetitions.
It shows  a \textit{learning process} along time, which stabilizes after a few repetitions.
On the one hand, all subjects gave the same ratings for at least half of the sequences starting from the third repetition. There were a few outliers, but only as expected of the distribution and these are not unduly influential.
On the other hand, all subjects changed their vote on at least 10\% of the sequences even at the 10th repetition.

\begin{figure*}[t]
\begin{tabular}{ccc}
\multicolumn{3}{c}{\includegraphics[width=0.5\linewidth]{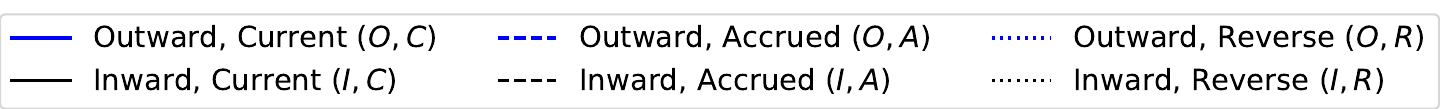}} \\
\includegraphics[width=0.3\linewidth]{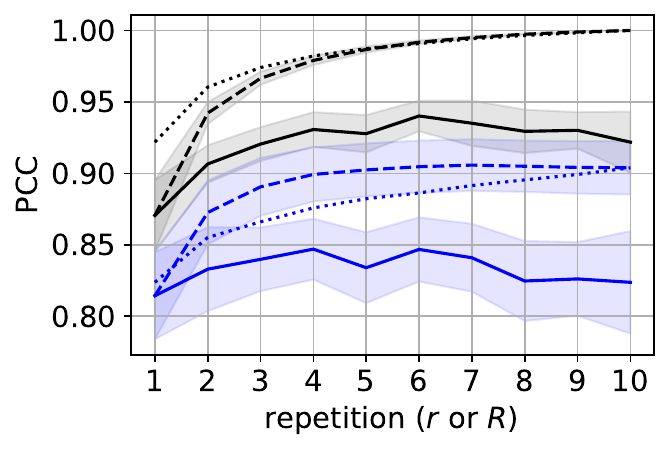}  & \includegraphics[width=0.3\linewidth]{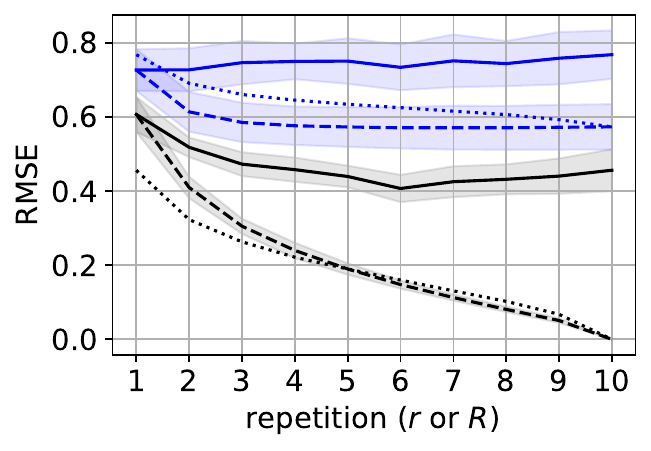}  & \includegraphics[width=0.3\linewidth]{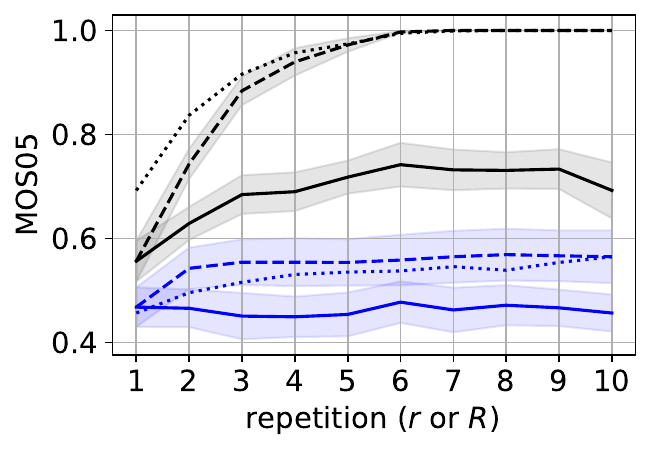} 
\end{tabular}
\caption{\label{fig:user_corr} 
Pearson correlation (left), Root Mean Square Error (center), and MOS05 (right) between individual and pooled ITERO scores, for four treatments of the ITERO subjective ratings.
Black lines show comparisons between each subject and his/her own estimated \textit{true opinion} $\hat{u}_{i,j}$ (inward), while blue lines show comparisons with the \textit{baseline} ITERO-ONE (outward).
Solid lines show results for \textit{current} individual repetition $r$, dashed lines \textit{accrue} the first $R$ repetitions, and dotted lines accrue the \textit{last} $R$ repetitions (\textit{reverse}).
}
\end{figure*}

It is interesting to see how this learning process results in information gain. 
To do so, we have compared the scores of the subjects in ITERO-TEN $u_{i,j,r}$ with the \textit{baseline} MOS from ITERO-ONE $\xi_j$ (\textit{outward} comparison), 
as well as with the mean opinion of the subject across the ten repetitions $\hat{u}_{i,j}$ (\textit{inward} comparison):

\begin{equation}
\label{eq:true_op_estimate}
\hat{u}_{i,j} = \frac{1}{10} \sum_{r=1}^{10} u_{i,j,r}
\end{equation}
which can be interpreted as an approximation of the subject \textit{true opinion} $\bar{u}_{i,j}$.

Fig. \ref{fig:user_corr} shows the results of this comparison for PCC, RMSE and MOS05 metrics, for four treatments of the ITERO subjective ratings.
Black lines show comparisons between each subject and his/her own estimated \textit{true opinion} $\hat{u}_{i,j}$ (inward), while blue lines show comparisons with the \textit{baseline} ITERO-ONE (outward).
Solid lines show results for \textit{current} individual repetition $r$, dashed lines \textit{accrue} the first $R$ repetitions, and dotted lines accrue the \textit{last} $R$ repetitions (\textit{reverse}).

Taking the left graphic (Pearson's linear correlation coefficient) as an example,
the solid blue line shows PCC between a single repetition of a single subject and the \textit{baseline}, averaged over all ITERO-TEN subjects:

\begin{equation}
\textsc{PCC}_{O,C}(r) = \frac{1}{20}\sum_{i=1}^{20} {\rho_j(u_{i,j,r}, \xi_j)}
\end{equation}
where $u_{i,j,r}$ represents an individual rating of the ITERO-TEN set, $\xi_j$ represents the \textit{baseline} MOS  of each PVS $j$ calculated from ITERO-ONE, $\rho_j(\cdot, \cdot)$ represents the correlation coefficient across all PVSs, and subscript (O,C) is an abbreviation for (Outward, Current).

The dashed blue line shows the correlation between the first $R$ repetitions of a single subject and the \textit{baseline}, averaged over all ITERO-TEN subjects:
\begin{equation}
\textsc{PCC}_{O,A}(R) = \frac{1}{20}\sum_{i=1}^{20} {\rho_j \left(\frac{1}{R} \sum_{r=1}^R u_{i,j,r}, \xi_j \right)}
\end{equation}
where subscript (O,A) is an abbreviation for (Outward, Accrued).
That is, the dashed blue line \textit{accumulates} or \textit{accrues} data from all prior repetitions, while the solid line considers only the \textit{current} repetition.  

The dotted blue line, labeled \textit{reverse}, considers the correlation between the \textit{last} R sessions of a single subject and the \textit{baseline}, averaged over all ITERO-TEN subjects:
\begin{equation}
\textsc{PCC}_{O,R}(R) = \frac{1}{20}\sum_{i=1}^{20} {\rho_j \left(\frac{1}{R} \sum_{r=10-R+1}^{10} u_{i,j,r}, \xi_j \right)}
\end{equation}
where subscript (O,R) is an abbreviation for (Outward, Reverse).

The black lines repeat these same computations, but they use, instead of \textit{baseline} ($\xi_j$), the \textit{inward} opinion of each subject ($\hat{u}_{i,j}$, see equation (\ref{eq:true_op_estimate})) across the ten repetitions:

\begin{align}
\textsc{PCC}_{I, C}(r) &= \frac{1}{20}\sum_{i=1}^{20} {\rho_j \left(u_{i,j,r},  \hat{u}_{i,j} \right)} \\
\textsc{PCC}_{I, A}(R) &= \frac{1}{20}\sum_{i=1}^{20} {\rho_j \left(\frac{1}{R} \sum_{r=1}^R u_{i,j,r}, \hat{u}_{i,j}  \right)} \\
\textsc{PCC}_{I,R}(R) &= \frac{1}{20}\sum_{i=1}^{20} {\rho_j \left(\frac{1}{R} \sum_{r=10-R+1}^{10} u_{i,j,r}, \hat{u}_{i,j} \right)}
\end{align}
where subscripts (I,C), (I,A), and (I,R) are an abbreviation for (Inward, Current), (Inward, Accrued), and (Inward, Reverse) respectively. 

The shaded area around each line represents the 95\% confidence interval of the mean. 
Confidence intervals have been excluded from \textit{reverse} plots for figure clarity, but they are similar to the ones for their respective \textit{accrued} plots.
The same treatments have been applied to RMSE (Fig. \ref{fig:user_corr}, center) and MOS05 (right).


\subsection{Inward comparison}
As already observed in Fig.~\ref{fig:user_changes}, Fig.~\ref{fig:user_corr} shows a  \textit{learning process} of the subjects with respect to their final opinions, as the first session tends to be farther away from the inward MOS than the rest. 
This learning process stops at about the fourth or fifth repetition; even when the inward accrued correlation, RMSE, and MOS05 converge to the final opinion of the subject, as expected, each individual repetition does not.

Individual scores $u_{i,j,r}$ at each repetition $r$ are, by definition, integer, and therefore they present some \textit{quantification noise} when used to approximate the estimated \textit{true opinion} of each subject $\hat{u}_{i,j}$.
Accruing the results of several repetitions can remove this noise and, after the sixth repetition, subjects have converged to their estimated \textit{true opinions} within our target resolution of 0.5 MOS points, i.e. $MOS05_{I,A}(6) = 1$.

\subsection{Outward comparison}

Each individual repetition has similar properties of outward \textit{current} association ($\textsc{PCC}_{O,C} \approx 0.83$), agreement ($\textsc{RMSE}_{O,C} \approx 0.75$) and perceptual similarity ($\textsc{MOS05}_{O,C} \approx 0.45$).
Unlike inward comparisons, the learning process described above does not result in each individual session being closer to the \textit{baseline} than the previous one.

However, this learning process produces some additional information which makes outward accrued metrics to actually converge to the \textit{baseline}.
Part of this convergence may be just the compensation for quantification noise.
But another part is truly produced by the learning process: the changes in opinion of the subject during the first repetitions actually generate information.
This can be seen when comparing (direct) accrued to reverse curves: the former converges to a saturation point much faster than the latter.

The behaviors described above apply similarly to association, agreement, and perceptual similarity.
After the first four or five repetitions, the average subject has $\textsc{PCC}_{O,A} \approx 0.9$, $\textsc{RMSE}_{O,A} \approx 0.6$, and $\textsc{MOS05}_{O,A} \approx 0.55$.
No significant improvement is produced after that, which can be interpreted as the opinion changes being just random noise.

\subsection{Subject bias}

Bias has been computed with respect to the \textit{baseline} (ITERO-ONE); and the distributions are calculated for the 20 subjects who completed all repetitions:
\begin{equation}
\delta_{i,j,r}   =  u_{i,j,r}-\xi_j\\
\end{equation}
\begin{equation}
\Delta_{i,r} = \frac{1}{110} \sum_{j} \delta_{i,j,r}  
\end{equation}

\begin{figure}[htb]
\centering
\includegraphics[width=0.95\columnwidth]{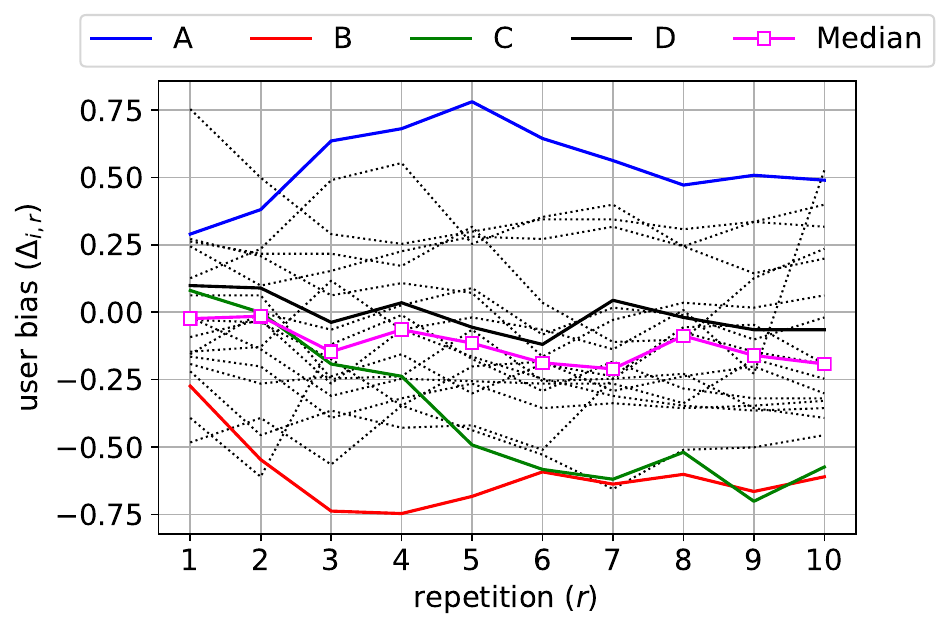} 
\caption{\label{fig:user_bias} Distribution of subject biases, where the bias removed is the \textit{baseline} bias of the first repetition. This estimates the overall shift of ratings as repetition increases. }
\end{figure}

Fig. \ref{fig:user_bias} shows the the bias of the different subjects for each repetition $\Delta_{i,r}$, as well as the median.
The subjects tend to get slightly more pessimistic with time, as seen in the median, but the distribution is relatively stable otherwise.
Four individual subjects have been identified to illustrate different behaviors: 
A and B are consistently \textit{optimistic} or \textit{pessimistic}.
C is the subject whose bias has highest variance; it transits from 0.1 to -0.6 as sequences advance.
D is the subject whose bias has lowest variance. It is relatively stable; however, it shows oscillations.

From Fig.~\ref{fig:user_bias} we can see that bias is not perfectly stable. To better understand its instability we tested whether bias for one session is statistically different from bias obtained for all sessions.
For all repetitions we can calculate mean bias by:

\begin{equation}
c_{i,j} = \frac{1}{10}\sum_{r=1}^{10}\delta_{i,j,r}  
\end{equation}

We have compared each subject $i$ and each repetition $r$ using the Student's \textit{t}-test $\delta_{i,j,r}$ to $c_{i,j}$, to determine whether global bias for subject $i$ is statistically different from that obtained for his/her session $r$.
Since we run multiple comparisons, we use Bonferroni significant level correction \cite{brunnstrom2018statistical}. 
The analysis shows that for 8 subjects we have no statistically different sessions, for 6 we have one statistically different session, for 5 two, and for 1 three statistically different sessions. 
The obtained results are tricky to interpret. For most comparisons we do not detect a statistically significant difference, but this is not always true. 

Additionally, considering the bias values obtained for subsequent repetitions, a slow global decreasing trend can be observed.
Such a trend cannot be related to bias instability, since it is a systematic change. 
We do not know why such a bias trend is observed. People seem to become slightly more critical as they see the same videos repeatedly. 
This increases the likelihood of detecting statistical significance between different repetitions. 
Considering all this, we think that the bias is mostly stable.

\begin{figure}[htb]
\centering
\includegraphics[width=0.95\columnwidth]{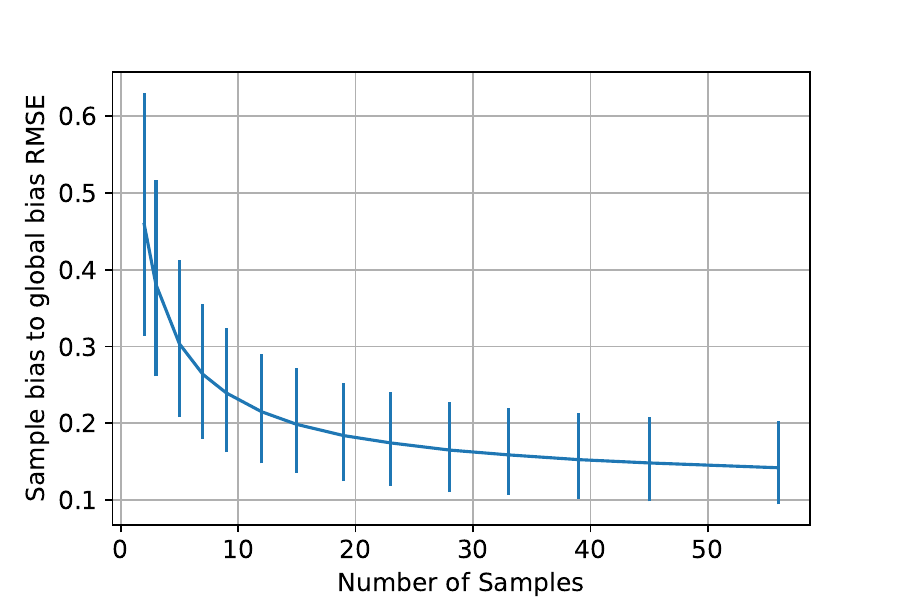} 
\caption{\label{fig:biasPred} The change of the root mean square error between the bias estimated based on all samples compared with the bias estimated based only on the $n$ samples.}
\end{figure}

A complementary analysis is studying how many sequences are needed to estimate bias from a complete experiment.
In this line, Fig.~\ref{fig:biasPred} shows the root mean square error of the global bias versus bias predicted by $n$ samples.
To compute it, we randomly chose a session and then randomly chosen $n$ samples out of the 110 sequences of the session.
Results suggest that bias can be predicted with around 15 samples with error around 0.2; for 60 samples the bias estimation error is around 0.15. 

In summary, our results suggest that subject bias actually exists and it is stable across sessions.
However, there is always going be an error (of at least around 0.15) when estimating the bias from specific sessions; 
and this error will be higher if the estimation is done from a reduced subset of sequences.


\section{Results for several observers}
\subsection{Estimating the baseline}

So far we have seen that a single subject has limited ability to predict the results of a ``traditional'' experiment such as the \textit{baseline}.
After approximately four repetitions, there is no visible gain of information or prediction capability.
We will now analyze whether this limitation can be overcome by aggregating the results of a reduced number of subjects.

To do so, we pick $S_N$, a random sample of $N$ subjects from ITERO-TEN, and we consider the MOS obtained by their first $R$ repetitions $\mu_{j,R}$.
We then compute our benchmark metrics between each subset and a modified baseline $\xi'_j$ that excludes the ratings from $S_N$ selected subjects\footnote{Note that $\mu_{j, R}$ and $\xi'_{j}$ depend on $S_N$, but we have decided not to show it explicitly to simplify the notation.}:

\begin{align}
\mu_{j,R} &= \frac{1}{N} \frac{1}{R} \sum_{i \in S_N} \sum_{r=1}^{R} u_{i,j,r} &u_{i,j,r} \in  \textsc{ITERO-TEN} \\
\xi'_{j} & = \frac{1}{27-N} \sum_{i \notin S_N}  u_{i,j} &u_{i,j} \in  \textsc{ITERO-ONE}
\end{align}

For instance, Pearson correlation is computed as
\begin{equation}
\label{eq:pcc_baseline}
\textsc{PCC}_{S_N}(R) = \rho_j(\mu_{j,R}, \xi'_{j} ) 
\end{equation}
where $\rho_j(\cdot, \cdot)$ represents the correlation coefficient across all PVSs. RMSE and MOS05 are computed likewise.

For each combination of $R$ and $N$, we repeat the process 1000 times and then compute the distribution of the benchmark metrics.

\begin{figure*}[ht]
\begin{tabular}{ccc}
\multicolumn{3}{c}{\includegraphics[width=0.65\linewidth]{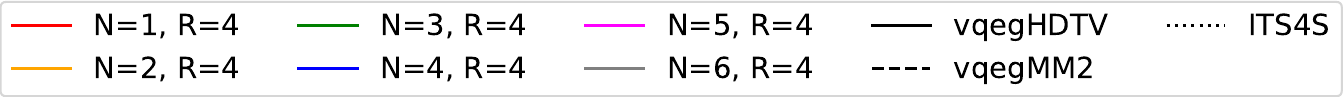}} \\
\includegraphics[width=0.3\linewidth]{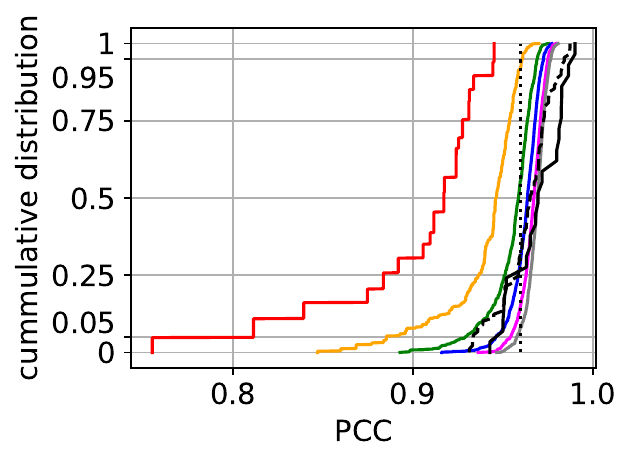} & \includegraphics[width=0.3\linewidth]{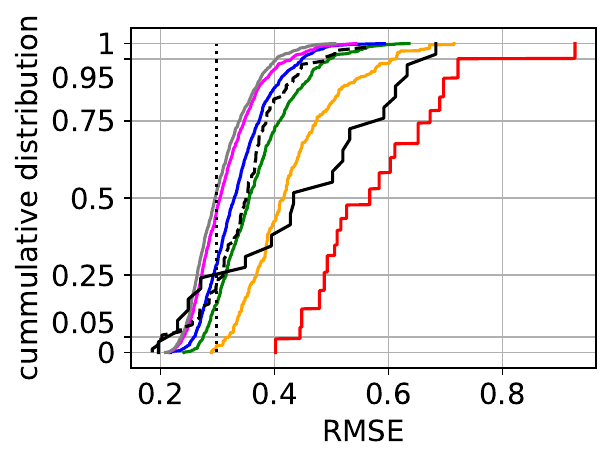}  & \includegraphics[width=0.3\linewidth]{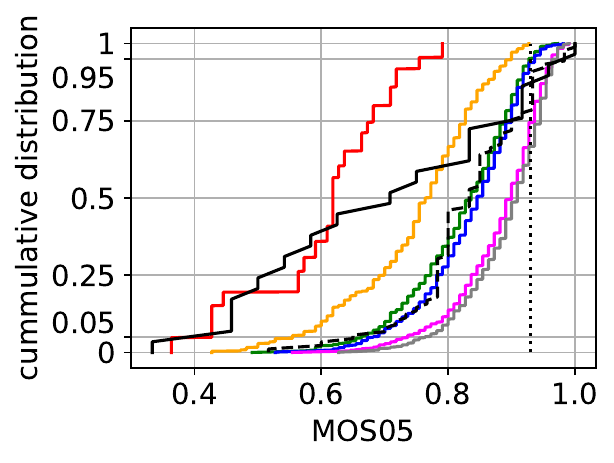}  \\
\includegraphics[width=0.3\linewidth]{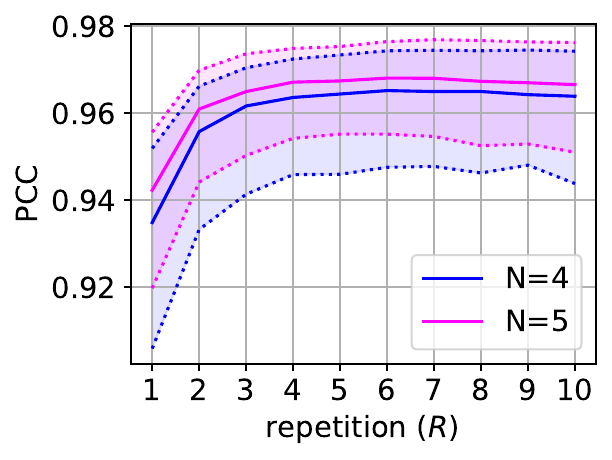} & \includegraphics[width=0.3\linewidth]{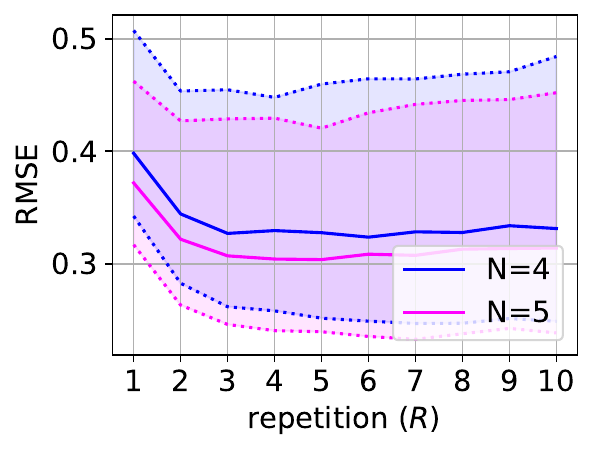}  & \includegraphics[width=0.3\linewidth]{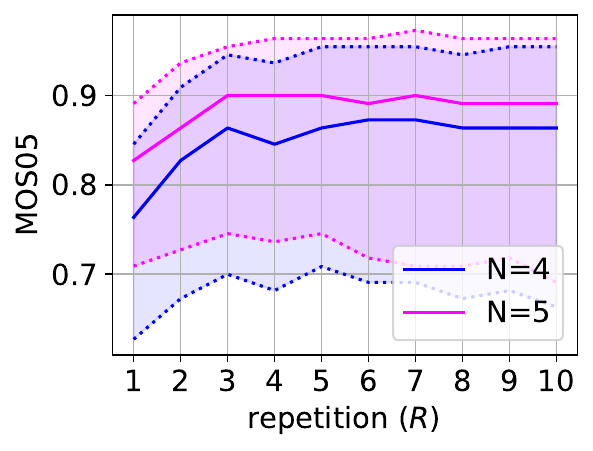} 
\end{tabular}
\caption{\label{fig:metrics_hist_4} Comparing ITERO-TEN with ITERO-ONE (\textit{baseline}). The top line shows cumulative distribution of the benchmark metrics for different numbers of subjects (1 to 6) and 4 repetitions. The distribution of pairwise comparisons between different labs in other subjective experiments (VQEG-HDTV, VQEG-MM2, ITS4S) is shown for reference. The bottom line shows the evolution of the metrics with repetition, for 4 and 5 subjects. Solid lines show the median value and dotted lines show 5th and 95th percentiles.}
\end{figure*}

\begin{table}[htb]
\centering
\caption{Comparing with baseline}
\label{tab:metrics_hist_4}
\begin{tabular}{|ccc|ccc|}
\hline
Perc & N & R & PCC & RMSE & MOS05 \\
\hline
Median & 1 & 4 & 0.917 & 0.567  & 0.618 \\
Median & 2 & 4 & 0.946 & 0.416 & 0.764 \\
Median & 3 & 4 & 0.959 & 0.355 & 0.835 \\
Median & 4 & 4 & 0.964 & 0.330 & 0.845 \\
Median & 5 & 4 & 0.967 & 0.304 & 0.900 \\
Median & 6 & 4 & 0.969 & 0.296 & 0.909 \\
\hline
.05/.95 & 1 & 4 & 0.811 & 0.722  & 0.427 \\
.05/.95 & 2 & 4 & 0.885 & 0.613 & 0.555 \\
.05/.95 & 3 & 4 & 0.929 & 0.496 & 0.664 \\
.05/.95 & 4 & 4 & 0.946 & 0.448 & 0.682 \\
.05/.95 & 5 & 4 & 0.954 & 0.429  & 0.736 \\
.05/.95 & 6 & 4 & 0.958 & 0.403 & 0.764 \\
\hline
\end{tabular}
\end{table}

The first row of
Fig.~\ref{fig:metrics_hist_4} shows the cumulative distribution of  the metrics for $1 \le N \le 6$ and $R=4$. 
Table~\ref{tab:metrics_hist_4} shows the values of the most relevant points of the distribution: median and 5th (PCC, MOSO5) or 95th (RMSE) percentiles.
We have initially selected $R=4$ as it is the point where additional repetitions do not seem to increase information from individual subjects, as described above.
As a reference, we have also computed the distribution metrics resulting from pairwise comparing different repetitions of the same subjective experiment, as also described in section \ref{sec:benchmark} (see Table \ref{tab:subjective-compare}): VQEG-HDTV (6 laboratories, 15 comparisons), VQEG-MM2 (10 laboratories, 45 comparisons) and ITS4S (2 laboratories, 1 comparison).

By most metrics, having 4 or 5 subjects repeat the session 4 times each outperforms the results of predicting one ``standard" experiment from the results of another one, both considering the median and the worst case (95\% percentile).
Having 4 subjects repeat the assessment 4 times results in correlation coefficients higher than 0.94, RMSE values lower than 0.45, and MOS05 higher than 0.68, with a probability higher than 95\%, also in line with state-of-the-art objective metrics (see Table \ref{tab:objective-vqeghd3}).

The second row of Fig.~\ref{fig:metrics_hist_4} shows the evolution of the metrics with respect to the repetitions.
The most relevant points of the cumulative distribution are shown: median, and edge percentiles (5\% and 95\%).
It can be seen that the behavior is similar to that of a single observer: it clearly improves during the first 3 repetitions, and it stabilizes at the 4th or 5th.
The ability to improve during repetition seems to be slightly better in association (correlation) than in agreement (RMSE) or perceptual similarity (MOS05), particularly for the worst case.


This limitation can be explained by the combined bias of the users within the same subset:
\begin{equation}
\Delta_{S_N}(R) = \frac{1}{110} \sum_{j} \left(  \mu_{j,R} - \xi'_{j}\right)
\end{equation}

According to the subject bias model in \eqref{eqn:bias}, $\Delta_{S_N}$ is the sum of $N$ independent normal random variables $\mathcal{N}(0, \sigma_{\Delta})$: 
\begin{align}
\Delta_{S_N}(R) &\sim \mathcal{N}(0, \sigma_{N,R}) \\
\sigma_{N,R} &\approx \frac{1}{\sqrt{N}}\sigma_{\Delta}, \enspace \forall R \label{eq:sigma_n}
\end{align}

The actual distribution of $\Delta_{S_N}$ is approximately Gaussian, with observed mean $\bar{\mu}_{N,R}$ and standard deviation $s_{N,R}$ shown in Table \ref{tab:bias_stats}.
It approximately follows \eqref{eq:sigma_n}.
Therefore, subject bias imposes a practical limitation on the achievable agreement between an experiment with a reduced number of subjects and a ``traditional'' one:
with only a few subjects, the combined bias will result in  systematic error affecting RMSE and MOS05 metrics.
In the next subsection we will explore whether it is possible to remove such bias by introducing some additional sequences in the experiment for that purpose (``Sports'' sequences, in our case).

\begin{table}[htb]
\centering
\caption{Distribution of bias}
\label{tab:bias_stats}


\resizebox{\columnwidth}{!}{
\begin{tabular}{|c|cccccc|ccc|}
\hline
N & 1 & 2 & 3 & 4 & 5 & 6 & 4 & 4 & 4 \\
R & 4 & 4 & 4 & 4 & 4 & 4 & 1 & 7 & 10 \\
\hline
$\bar{\mu}_{_{N,R}}$ & -.06 & -.06 & -.05 & -.05 & -.05 & -.05 & -.00 & -.07 & -.09 \\ 
$s_{_{N,R}}$ & .29 & .21 & .18 & .16 & .15 & .12 & .15 & .16 & .16 \\
\hline
\end{tabular}
}
\end{table}

\subsection{Estimating the  ground truth}

\begin{figure*}[ht]
\begin{tabular}{ccc}
\multicolumn{3}{c}{\includegraphics[width=0.65\linewidth]{figs/cumulative_legend.pdf}} \\
\includegraphics[width=0.3\linewidth]{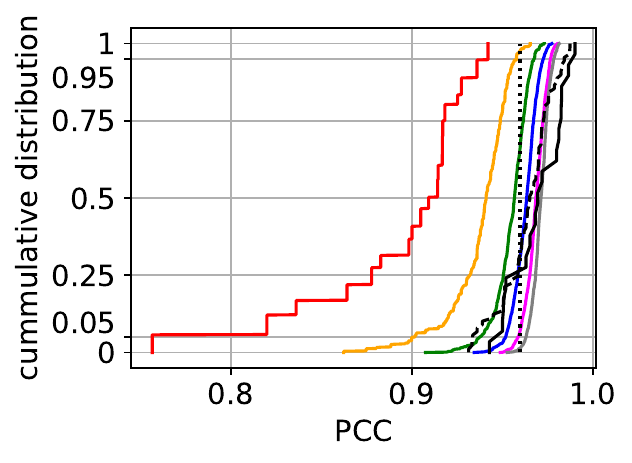} & \includegraphics[width=0.3\linewidth]{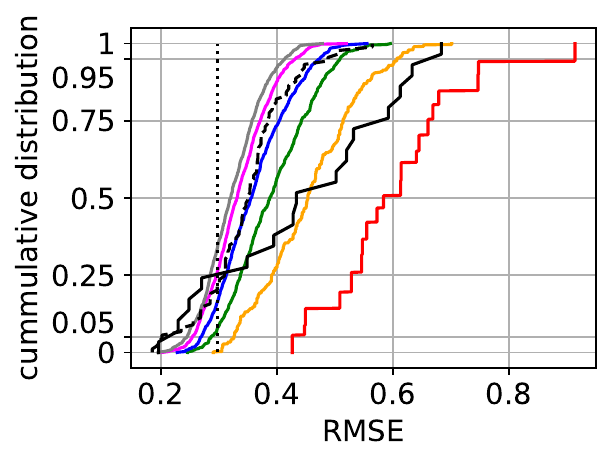} & \includegraphics[width=0.3\linewidth]{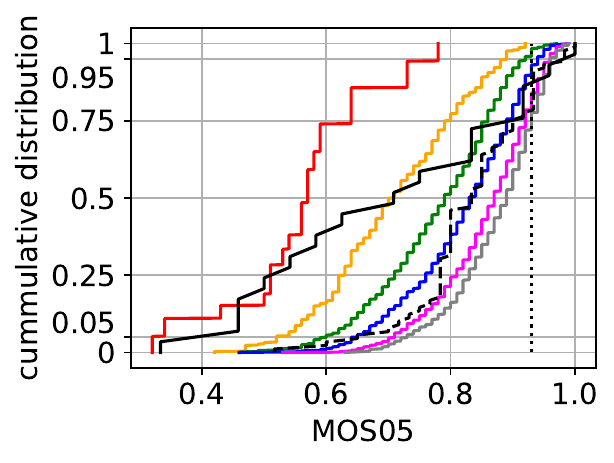}   \\
\includegraphics[width=0.3\linewidth]{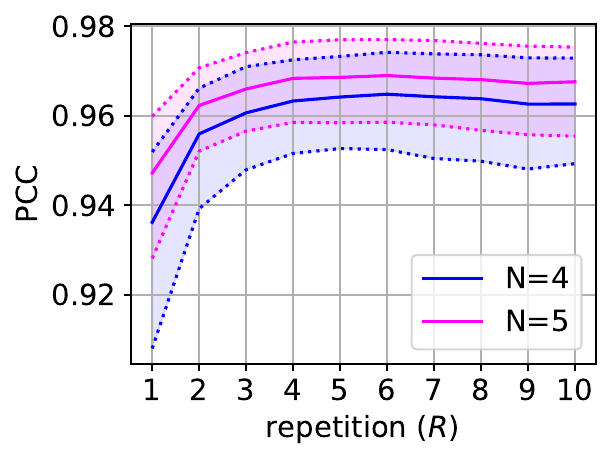} & \includegraphics[width=0.3\linewidth]{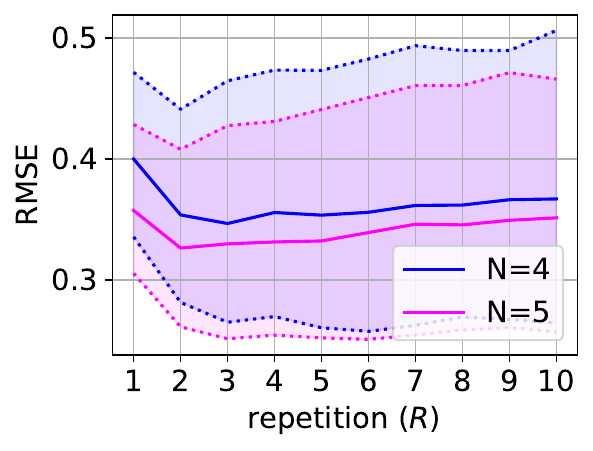}  & \includegraphics[width=0.3\linewidth]{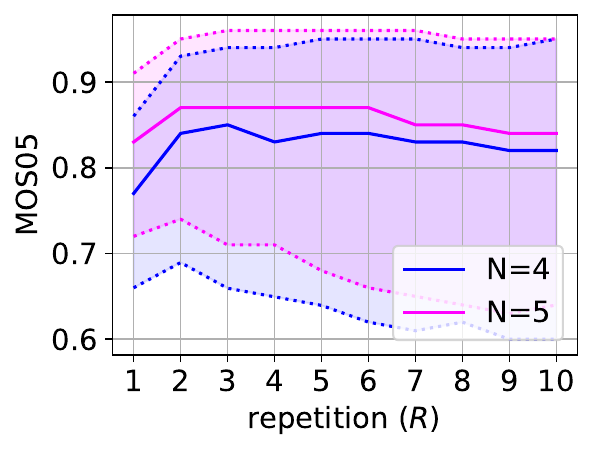}  \\
\end{tabular}
\caption{\label{fig:metrics_its4s} Comparison of the metrics with a ground truth of the original ITS4S experiment (``Everglades'' sequences only). The same analysis as in Fig.~\ref{fig:metrics_hist_4} has been used.}
\end{figure*}

So far we have shown the performance of the metrics of $N$ subjects, $R$ repetitions, when predicting the values of the very same experiment done by  $27-N$ subjects, once (\textit{baseline}).
However, the final target would be to be able to predict the original ``Everglades'' results in ITS4S database (\textit{ground truth}).
In fact, the addition of the ``Sports'' sequences should be helpful to estimate and correct the combined bias of the subjects.

The same analysis described in the previous subsection is done to compute the benchmark metrics of a random subset of ITERO-TEN with the ground truth ITS4S.
Now \eqref{eq:pcc_baseline} is replaced by:
\begin{equation}
\label{eq:pcc_groundtruth}
\textsc{PCC}_{S_N}(R) = \rho_j(\mu_{j,R}, \hat{\psi}_{j} ) 
\end{equation}
and likewise for the rest of the metrics. 
$\hat{\psi}_{j}$ is the MOS of each PVS in the original ITS4S ``Everglades'' experiment, and $\rho_j$ is computed only for the 100 ``Everglades'' sequences.

Fig.~\ref{fig:metrics_its4s} shows some of the results.
Results are similar in terms of \textit{association}, but the behavior is worse in terms of \textit{agreement} or \textit{perceptual similarity}; even results do not differ very much, it is clear that the addition of repetitions does not improve RMSE or MOS05 significantly.

\begin{figure*}[ht]
\begin{tabular}{ccc}
\includegraphics[width=0.28\linewidth]{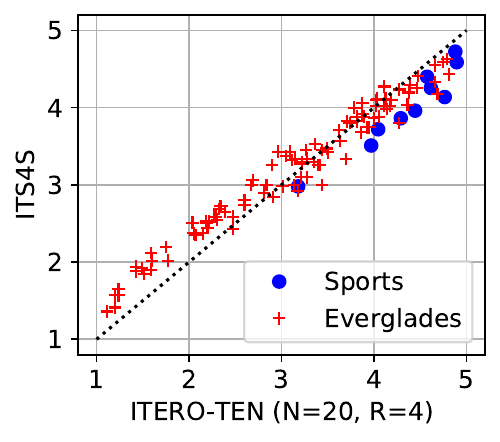} & \includegraphics[width=0.35\linewidth]{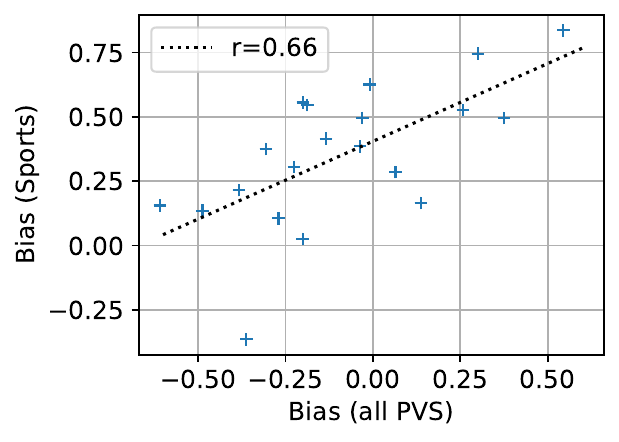}  & \includegraphics[width=0.3\linewidth]{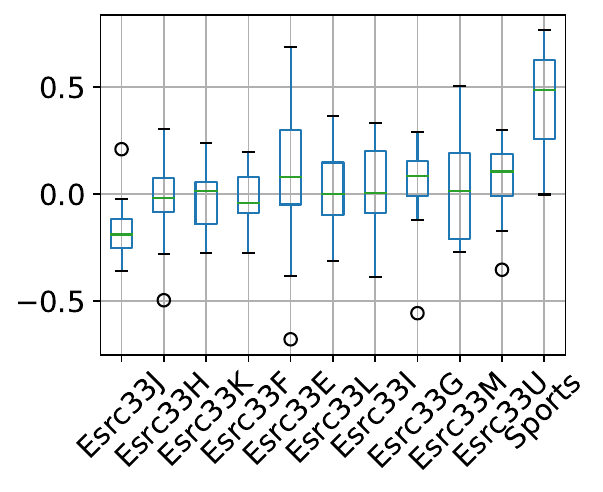} \\
\footnotesize(a) & \footnotesize(b) & \footnotesize(c) 
\end{tabular}
\caption{\label{fig:bias_groundruth} (a) Comparison of MOS values between ground truth ITS4S and ITERO-TEN experiment, showing global average difference for ``Everglades'' and ``Sports'' sequences. 
(b) Comparing individual subject bias in ``Sports'' sequences with their global bias for the all 110 PVSs. 
(c) Distribution of error in bias estimation when considering only a few sequences, for different selection of such sequences (all the different SRCs in ITS4S experiment, plus the added ``Sports'' sequences).}
\end{figure*}

A potential explanation for this is the fact that the introduction of the 10 ``Sports'' sequences within the ``Everglades'' dataset
actually modified the scoring scales of the subjects.
Fig.~\ref{fig:bias_groundruth}(a) shows the MOS of each individual sequence in ITS4S vs ITERO-TEN, considering all users and four repetitions ($N=20, R=4$).
Two different effects can be observed: on the one hand, the spread of the distribution is wider in the case of ITERO-TEN;
on the other, it is clear that  ``Everglades'' and ``Sports'' sequences are scored differently. 
On average, Everglades sequences are scored 0.09 points lower in ITERO-TEN than in the ITS4S.
However, ``Sports'' sequences are scored 0.35 points \textit{higher} in ITERO-TEN.

As a consequence, it is not possible to compensate for the subject bias in ''Everglades`` sequences with only the bias estimated from ``Sports'' sequences; it actually makes results worse (\textit{increases} RMSE and \textit{reduces} MOS05).
Fig.~\ref{fig:bias_groundruth}(b) shows that, even though there is some correlation between predicted and actual bias, it is not strong enough to really improve the agreement properties of the MOS calculation. 
This is in line with Fig.~\ref{fig:biasPred} showing that 10 sequences do not allow precise estimation of bias. 

A possible reason for the ``Sports'' sequence not being able to predict general subject bias is that the kind of content may be significantly different from ``Everglades''.
To explore this possibility, we have used different subsets of sequences (within the 110 PVS) to estimate the bias of the whole experiment.
In particular, we have taken each of the SRCs, as defined in ITS4S dataset, as the bias estimator.
Unlike other subjective experiments, different PVSs of the same ITS4S dataset contain \textit{different} source content; however, they belong to the same scene as the original content.
Fig.\ref{fig:bias_groundruth}(c) shows the distribution of bias estimation error for each SRC, including the ``Sports'' subset.
As can be seen, the error behavior depends quite significantly on the specific sequences selected to compute it, and therefore it is not stable across experiments.


\subsection{Confusion Analysis}
When a subjective test is repeated in multiple labs, each lab will reach slightly different conclusions. For all pairs of stimuli, A and B, we will use the paired stimulus Student’s \textit{t}-test to decide whether A is better than, equivalent to, or worse than B. We can then compare the conclusions reached by the different labs and tally their frequency. 

We are only interested in two of the outcomes. The first is the likelihood that the two subjective tests disagree (i.e., the labs reach opposing conclusions on the quality ranking of A and B). Pinson \cite{SubjObjCI} shows that the disagree rate is stable and not influenced by the number of subjects or range of quality in the subjective test. 
We expect a well-designed and carefully conducted subjective test to have a disagree rate $\le$ 1\%. 

The second outcome of interest is the likelihood that two subjective test labs agree (i.e., the labs reach the same conclusion on the quality ranking of A and B, ignoring ties). The agree rate is influenced by the number of subjects and range of quality in the subjective test, so we must compare results with statistics gathered from ITERO-ONE and the ground truth. This gives us three labs, each with 24 or 27 subjects. Using all available subjects and the Everglades PVSs, lab-to-lab comparisons yield agree rates of 66\%, 66\%, and 68\%. If we randomly select 15 subjects, the agree rate ranges from 52\% to 63\%, with an average of 57\%.

We will use the 100 Everglades sequences and the ITERO-TEN subjects to compute agree and disagree rates for different numbers of subjects and repetitions. For each case, we will randomly select subjects and repetitions, and then compare those ratings to the ground truth data from each of the ground truth labs separately. This random selection will be repeated 50 times, for a total of 100 trials. 

Based on this data, Table \ref{tab:15subj} shows the likelihood in percent that a test with \textbf{N} subjects and \textbf{R} repetitions will have rates of agreement ($\ge$ 52\%) and disagreement ($\le$ 1\%) equivalent to a conventional subjective test of 15 subjects.
Table \ref{tab:24subj} repeats this analysis for a 24 subject test, where equivalence requires agreement $\ge$ 66\% and disagreement $\le$ 1\%.
Each column contains a single number of subjects (e.g., ``1S'' means one subject, ``4S'' means four subjects).

From Table \ref{tab:15subj} and \ref{tab:24subj}, let us choose experiment designs where the likelihood of equivalence is $\ge$ 95\%. We will add one repetition beyond minimum as a safety margin, because our statistics unrealistically assume that no other factors will cause the disagree rate to rise.
These criteria, combined with our desire for a small number of subjects, identify the following experiment designs for 15 subjects:
\begin{itemize}
    \item 3 subjects \& 5 repetitions $\approx$ 15 subjects
    \item 4 subjects \& 4 repetitions $\approx$ 15 subjects
    \item 5 subjects \& 3 repetitions $\approx$ 15 subjects
\end{itemize}
And these experiment designs for 24 subjects:
\begin{itemize}
    \item 5 subjects \& 6 repetitions  $\approx$ 24 subjects
    \item 6 subjects \& 5 repetitions  $\approx$ 24 subjects
\end{itemize}
The lower values in \ref{tab:24subj} reinforces the theory that the 24 subject test is a higher standard of performance than the 15 subject test. 

\begin{table}[htb]
\centering
\caption{Likelihood Equivalent to 15 Subject Test} 
\label{tab:15subj}
\begin{tabular}{ c | c c c c c c c c}
 reps & 1S & 2S & 3S & 4S & 5S & 6S & 7S & 8S \\ \hline
    1 & na & 0  & 0  & 0  & 1  & 59 & 90 & 100 \\
    2 &  0 & 0  & 32 & 94 & 98 & 99 & 99 & 100 \\
    3 &  0 & 16 & 89 & 97 & 99 & 100 & 99 & 100 \\
    4 &  0 & 55 & 96 & 97 & 98 & 98 & 99 & 100 \\
    5 &  0 & 70 & 95 & 98 & 98 & 99 & 99 & 100 \\
    6 &  3 & 85 & 93 & 97 & 98 & 99 & 99 & 98 \\
    7 &  3 & 77 & 93 & 95 & 99 & 96 & 98 & 99 \\
    8 &  8 & 78 & 90 & 94 & 96 & 98 & 98 & 99 \\
    9 &  6 & 71 & 82 & 95 & 94 & 97 & 96 & 97 \\
   10 &  4 & 66 & 82 & 91 & 89 & 97 & 99 & 98 \\
\end{tabular}
\end{table}

\begin{table}[htb]
\centering
\caption{Likelihood Equivalent to 24 Subject Test} 
\label{tab:24subj}
\begin{tabular}{ c | c c c c c c c c}
 reps & 1S & 2S & 3S & 4S & 5S & 6S & 7S & 8S \\ \hline
    1 & na & 0  & 0  & 0  & 0  & 0  & 0  & 0  \\
    2 &  0 & 0  & 0  & 0  & 0  & 0  & 26 & 64 \\
    3 &  0 & 0  & 5  & 8  & 46 & 87 & 97 & 100 \\
    4 &  0 & 0  & 5  & 56 & 92 & 96 & 99 & 100 \\
    5 &  0 & 1  & 30 & 81 & 97 & 99 & 99 & 100 \\
    6 &  0 & 14 & 58 & 91 & 98 & 99 & 99 & 98 \\
    7 &  0 & 22 & 84 & 95 & 99 & 96 & 98 & 99 \\
    8 &  0 & 33 & 81 & 93 & 96 & 98 & 98 & 99 \\
    9 &  0 & 39 & 78 & 95 & 94 & 97 & 96 & 97  \\
   10 &  0 & 48 & 80 & 91 & 89 & 97 & 99 & 98 \\
\end{tabular}
\end{table}

\section{Discussion}

\subsection{FOWR test methodology}

After analyzing the results obtained by the ITERO experiment, we can state that the main research question has been answered affirmatively:
it is possible to use few observers with repetitions (FOWR) to obtain valid subjective scores, although with some limitations.

The experiment was designed to take place under conditions that are easy to replicate in any quality assessment laboratory: 
most subjects were actually staff of the laboratory, certainly including video experts, and the viewing conditions were not strict.
In fact, from the observers who completed the ten repetitions, only 4 of them can be considered ``naive subjects'' with respect to video quality.
They show similar results as the others, e.g. their combined performance for $R=4$ is PCC=0.96, RMSE=3.28, MOS05=0.89.
Therefore we can argue that our conclusions apply to \textit{expert viewers}, but they will probably be applicable to other kinds of observers as well.

To get good \textit{association} results, it is enough to do the experiment with 4 subjects and 4 repetitions.
Repetitions should be done on different days.
For \textit{agreement} and \textit{perceptual similarity}, however, there is a problem with combined bias.
Chances are good that 4 subjects are enough (median results still beat state-of-the-art metrics), but the distribution tails (5/95 percentiles) are worse.
For safer results, 5-6 subjects should be used.
In general, increasing the number of subjects will always improve the test result, while increasing the number of repetitions (beyond 4) will not.

Due to the inability to get an accurate agreement, the FOWR protocol cannot replace a full subjective assessment test.
However, it can provide good enough results for a \textit{pre-test}: to further prepare a subjective test.
It can also be used in the absence of an available objective score, with similar expected predictive capability. To put this into perspective, \cite{SubjObjCI} shows that some objective metrics perform equivalently to subjective tests of 24 subjects, when confidence intervals are used to make decisions.  

\subsection{On the limits of the subject model}

Subject model \eqref{eqn:subjectmodel} assumes that ratings in a subjective test follow a specific random process. 
Our experiment confirms this hypothesis by showing that even the same subject repeating the same experiment generates different answers.
These differences are beyond the scale limitation even for a very simple five point scale. 
Future analysis should focus on better understanding and limiting answer randomness. 

The experiment repetition provides interesting data about subject bias. 
First we see that we can estimate subject bias with limited precision, which is not surprising taking into account the typical precision of a psychological test. 
We see that for most subjects the bias is stable, so we can count it as an important model parameter. 
On the other hand, we showed that using different content to estimate bias does not work well. 
One explanation is that we need more sequences (see Fig.~\ref{fig:biasPred}): 10 sequences are not enough to estimate bias for a 100-sequence experiment.
Another is that the bias estimated with some sequences in one specific experiment cannot be used to predict the bias of those sequences within a different experiment.
Besides, introducing just 10\% additional sequences can affect the scores of the sequences under study.

Finally, we have also shown that there is a practical limitation on the ability to estimate the subject model parameters by repeating the same experiment several times.
Repetitions do not allow estimating the subject \textit{true opinion} $\bar{u}_{i,j}$ with precision, as would be expected from \eqref{eqn:true_opinion}.
Consecutive repetitions of random variable $U_{i,j,r}$ for a given subject and sequence are not independent.

\subsection{Implications for traditional experiments}

Even though there are limitations on the ability to compute the actual \textit{true opinion} $\bar{u}_{i,j}$, we have assumed that we can obtain a reasonable estimation $\hat{u}_{i,j}$ by averaging the results from all the available repetitions, as defined in \eqref{eq:true_op_estimate}.
Additionally, Fig.~\ref{fig:user_corr} shows that subject opinion converges after a few repetitions, so that the 10th repetition $u_{i,j,10}$ is closer to the estimated true opinion than the first one $u_{i,j,1}$.
However, ``traditional'' experiments (e.g. based on ITU-T P.910) only ask for the first opinion of each subject $u_{i,j,1}$.
Would results be different if the experiment was performed with a better estimate of $\bar{u}_{i,j}$, e.g. using several repetitions?

\begin{table}[htb]
\centering
\caption{Comparison of first and last repetition with per-subject mean value}
\label{tab:repetitions}
\begin{tabular}{|c|c|c|c|}
\hline
Comparison & PCC & RMSE & MOS05 \\
\hline
First vs Average & 0.990 & 0.176  & 0.991 \\
Last vs Average  & 0.995 & 0.101 &  1.000  \\
First vs Last    & 0.981 & 0.230 &  0.963 \\   
\hline
\end{tabular}
\end{table}

Fortunately, when considering the aggregate opinion of all the subjects, we have found no differences between the alternative estimates of $\bar{u}_{i,j}$.
Table~\ref{tab:repetitions} shows the pairwise comparison between the first repetition ($u_{i,j,1}$), the last one ($u_{i,j,10}$), and the average of all 10 sessions ($\hat{u}_{i,j}$), for the 20 subjects in ITERO-TEN subset:
coincidence is almost exact.
However, the distinction may be relevant when modeling the behavior of individual subjects.

Additionally, there are some implications for the design of large multi-site subjective tests, such as the ones performed by VQEG under HDTV or MM projects.
Those tests traditionally use different source contents in each laboratory.
However, there is typically a \textit{common set} which appears in all the tests, and is used to align the result across labs.
Unfortunately, introducing those extra sequences may alter the score of the sources under study in unexpected ways (see Fig.~\ref{fig:bias_groundruth}), particularly affecting agreement and perceptual equivalence of the experiments.

\section{Conclusions}

In this paper, we propose the FOWR experiment design, where a small number of subjects rate the same set of PVSs repeatedly, on different days. We prove that the FOWR experiment design is non-inferior to a conventional subjective test. 
By non-inferior, we mean the FOWR experiment design yields similar performance to a conventional experiment design, based on association, agreement, perceptual similarity, and confusion analysis.

We recommend the FOWR methodology for pilot studies (to indicate trending), for pre-tests, and as an alternative to objective metrics for laboratory applications. The FOWR experiment design is particularly valuable when an objective metric is not available (e.g., new technologies, camera capture). The FOWR method allows a small team to make a quick and reasonably accurate quality assessments, when the time and expense of subject recruitment is non-viable.  

For most applications, we recommend 4 subjects rating all stimuli 4 times on subsequent days. This experiment design is at least as good as the best objective metrics and will probably respond similarly to a 15 subject test.

There are intrinsic limitations on the protocol, particularly with respect to its capacity for agreement, as subject bias cannot be compensated.
If accurate agreement is required, we recommended 5 subjects scoring 5 times or 6 subjects scoring 5 times. These experiment designs will probably respond similarly to a 24 subject test.
We tested the FOWR protocol on expert subjects, who are more likely to be recruited for this type of in-house test. 

Subject bias exists and is reasonably stable across time.
However, subject bias is not uniform across sequences, either from the same or from different experiments.
There is ``behavioral correlation" (optimistic subjects tend to be more optimistic, average-wise).
However, subjects who are ``optimistic" with respect to one content sequence may be ``pessimistic" with respect to another.

Our results also have some implications for the modeling of subjective score processes.
First, the hypothesis that subject bias is independent of the PVS is only valid within a given subjective test; it is not stable across tests.
In addition, when adding some sequences from test A into experiment B, those sequences will not only be evaluated differently from how they were originally rated in A,
but also impact the evaluation of the sequences already present in B.
This challenges the whole concept of ``common set'' in cross-lab experiments.
And finally, when repeatedly rating the same set of sequences, subjects tend to converge to their \textit{true opinion} after about 4 repetitions.
When modeling the subject scoring process, this may conflict with traditional experimental design where subjects are instructed to rate PVSs that they have never seen before.

\section*{Acknowledgments}

The authors want to thank the subjects who volunteered to repeat the same subjective assessment experiment ten times.
 
The authors also want to thank Aleks Zale\'{n}ski (AGH) and Daniel Berj\'{o}n (UPM) for their help with the test setup.

\ifCLASSOPTIONcaptionsoff
  \newpage
\fi



\bibliographystyle{IEEEtran}
\bibliography{fowr}

%





\begin{IEEEbiography}[{\includegraphics[width=1in,height=1.25in,clip,keepaspectratio]{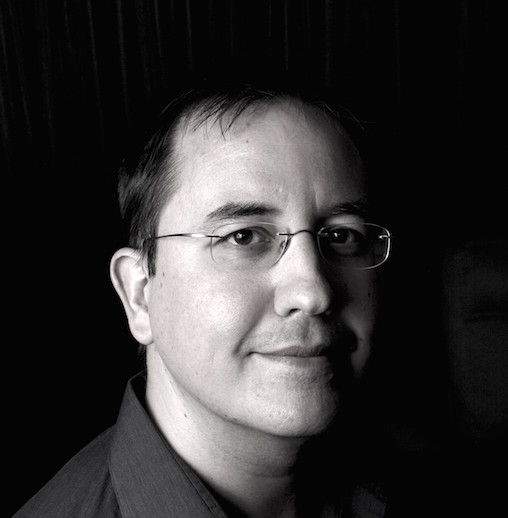}}]{Pablo Perez} received the Telecommunication Engineering degree (integrated BSc-MS) in 2004 and the Ph.D. degree in Telecommunication Engineering in 2013 (Doctoral Graduation Award), both from Universidad Politécnica de Madrid (UPM), Madrid, Spain. From 2004 to 2006 he was a Research Engineer in the Digital Platforms Television in Telefonica I+D and, from 2006 to 2017, he has worked in the R\&D department of the video business unit in Alcatel-Lucent (later acquired by Nokia), serving as technical lead of several video delivery products. Since 2017, he is Senior Researcher in the Distributed Reality Solutions department at Nokia Bell Labs. His research interests include multimedia quality of experience, video transport networks, and immersive communication systems.\end{IEEEbiography}

\begin{IEEEbiography}[{\includegraphics[width=1in,height=1.25in,clip,keepaspectratio]{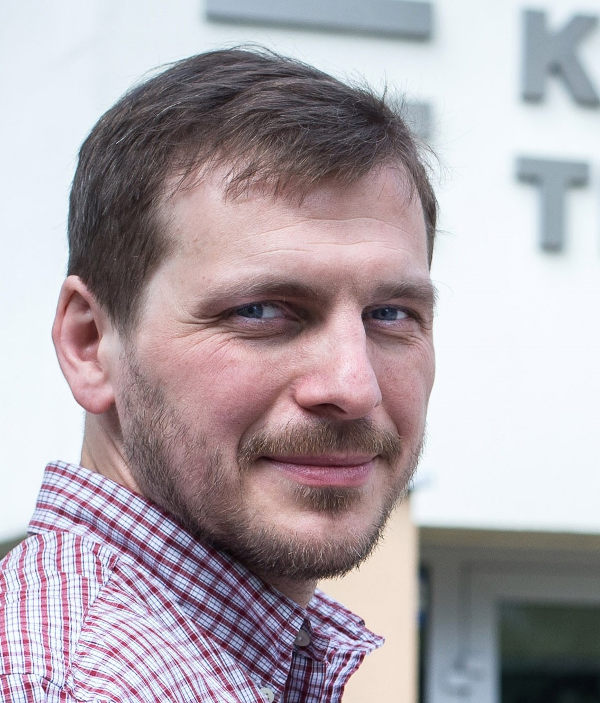}}]{Lucjan Janowski} received the Ph.D. degree in telecommunications from the AGH University of Science and Technology, Krakow, Poland, in 2006. In 2007, he was a Postdoctoral Researcher with the Laboratory for Analysis and Architecture of Systems, Centre National de la Recherche Scientifique, Paris, France. From 2010 to 2011, he was a Postdoctoral Researcher with the University of Geneva, Geneva, Switzerland. From 2014 to 2015, he was a Postdoctoral Researcher with The Telecommunications Research Center Vienna, Vienna, Austria. He is currently an AGH Professor with the Institute of Telecommunications, AGH University of Science and Technology. His research interests include statistics
and probabilistic modelling of subjective rates used in QoE evaluation.
\end{IEEEbiography}

\vfill
\newpage

\begin{IEEEbiography}[{\includegraphics[width=1in,height=1.25in,clip,keepaspectratio]{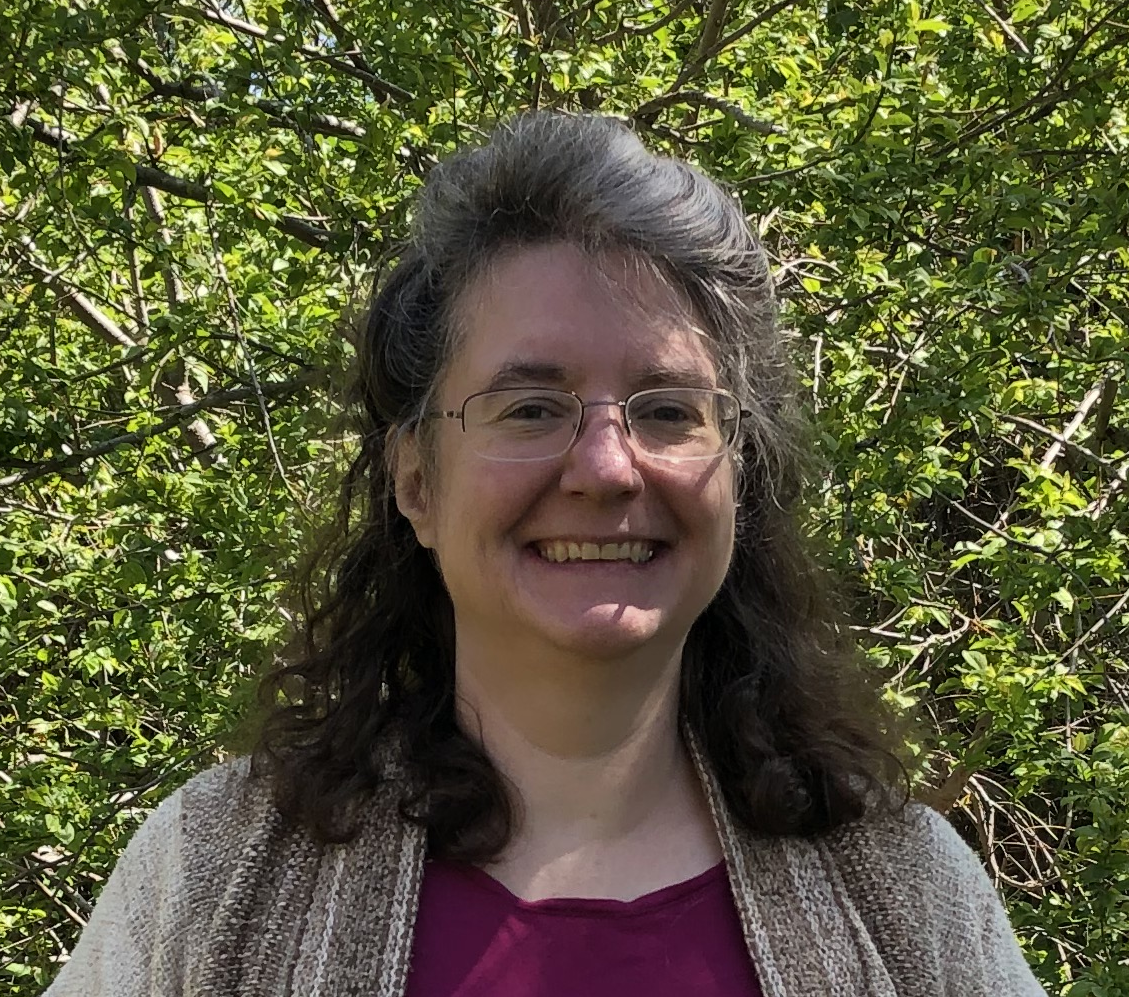}}]{Margaret Pinson} is an internationally recognized expert with 30 years of experience developing improved methods for assessing video quality. Her research includes algorithm development, human testing, and international standards. Her current research focuses on no reference (NR) metrics that predict what people would say is the quality of an image or video. NR metrics, when available, will enable smart cameras that adapt to first responder environments and applications. Mrs. Pinson is a Co-Chair of the Video Quality Experts Group (VQEG), administers the Consumer Digital Video Library (CDVL), and makes all of her algorithms openly available. Mrs. Pinson contributes to ITU Recommendations and has led several efforts to independently validate metrics, which is a necessary step of the standards development process. Mrs. Pinson designed and helped to conduct two prize challenges and has authored or co-authored 79 publications. 
\end{IEEEbiography}

\begin{IEEEbiography}[{\includegraphics[width=1in,height=1.25in,clip,keepaspectratio]{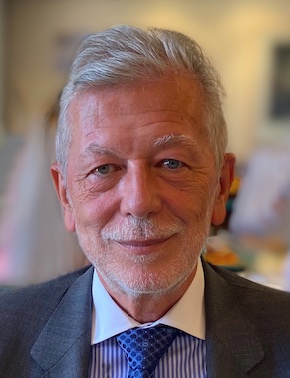}}]{Narciso García}received the Ingeniero de Telecomunicación degree (five
years engineering program) (Spanish National Graduation Award) and the
Doctor Ingeniero de Telecomunicación degree (Ph.D. in communications)
(Doctoral Graduation Award) from the Universidad Politécnica de Madrid
(UPM), Madrid, Spain, in 1976 and 1983, respectively. Since 1977, he has
been a member of the faculty of the UPM, where he is currently a
Professor of Signal Theory and Communications. He leads the Grupo de
Tratamiento de Imágenes (Image Processing Group), UPM. He has been
actively involved in Spanish and European research projects, also
serving as an Evaluator, a Reviewer, an Auditor, and an Observer of
several research and development programs of the European Union. He was
a Co-Writer of the EBU proposal and base of the ITU standard for digital
transmission of TV at 34–45 Mb/s (ITU-T J.81). He was an Area
Coordinator of the Spanish Evaluation Agency, from 1990 to 1992, and he
was the General Coordinator of the Spanish Commission for the Evaluation
of the Research Activity, from 2011 to 2014. He has been the Vice-Rector
for International Relations of the Universidad Politécnica de Madrid,
from 2014 to 2016. His current research interests include digital video
compression, computer vision, and quality of experience. He was a
recipient of the Junior and Senior Research Awards of the Universidad
Politécnica de Madrid in 1987 and 1994.
\end{IEEEbiography}

\vfill

\end{document}